\documentclass[aps,prx,superscriptaddress,twocolumn]{revtex4-2}
\usepackage{graphicx,color}
\usepackage{verbatim}
\usepackage{amssymb}   
\usepackage{amsmath}
\usepackage{amsfonts}
\usepackage{mathdots}
\usepackage{hyperref}
\usepackage{multirow}
\usepackage{epsfig}
\usepackage{hyperref}
\usepackage{xcolor}
\usepackage{cancel}
\usepackage{booktabs}
\usepackage{array}
\usepackage{array, makecell} 
\definecolor{myblue}{RGB}{46, 48,146}
\usepackage{multirow}
\hypersetup{colorlinks=true,linkcolor=myblue,citecolor=myblue,urlcolor=myblue, linktocpage}
\usepackage{braket}
\usepackage{bm}
\usepackage{multirow}

\begin{document}
	\title{  Quasi-one-dimensional  Supersolids in Luther-Emery Liquids}

	\author{Ying-Ming Xie}
	\affiliation{RIKEN Center for Emergent Matter Science (CEMS), Wako, Saitama 351-0198, Japan} 	
	\author{Naoto Nagaosa}\thanks{nagaosa@riken.jp}
	\affiliation{RIKEN Center for Emergent Matter Science (CEMS), Wako, Saitama 351-0198, Japan} 
	\affiliation{Fundamental Quantum Science Program, TRIP Headquarters, RIKEN, Wako 351-0198, Japan} 	
	
	\date{\today}
	\begin{abstract}
    
	The supersolid is a long-sought phase in condensed matter physics, characterized by the coexistence of density wave and superfluid orders. This phase is counterintuitive, as different symmetry-breaking orders typically compete with one another. A deeper understanding of how such a state forms in condensed matter systems remains an open question, especially in quasi-one-dimensional correlated systems. In this work, we investigate the emergence of supersolids in  Luttinger-Emery liquids using a variational method.  As the system consists of coupled Luttinger-Emery liquid chains, we refer to this phase as a quasi-one-dimensional supersolid. Notably, we demonstrate that the quasi-one-dimensional supersolid phase is energetically favorable in chains with finite size or short-range order. Furthermore, we investigate the collective dynamics of these coexisting CDW and SC states, identifying a quasi-Goldstone mode. Our theory provides valuable insights into both the ground state and the dynamic properties of supersolids in strongly correlated systems.
	
	\end{abstract}
	
	\maketitle
	
	\section{Introduction}

Strongly correlated materials often exhibit multiple broken-symmetry phases. A well-known example is the cuprates, which host charge density waves (CDW), spin density waves, and superconducting orders across a broad range of material parameters. Notably, evidence suggests that the interplay between various fluctuating or static order parameters plays a crucial role in shaping the pseudogap regime of high-temperature superconductors \cite{Fradkin2015,Agterberg}. More broadly, understanding how different symmetry-broken orders compete or intertwine represents one of the most fundamental problems in condensed matter physics.

\begin{figure}
    \centering
    \includegraphics[width=1\linewidth]{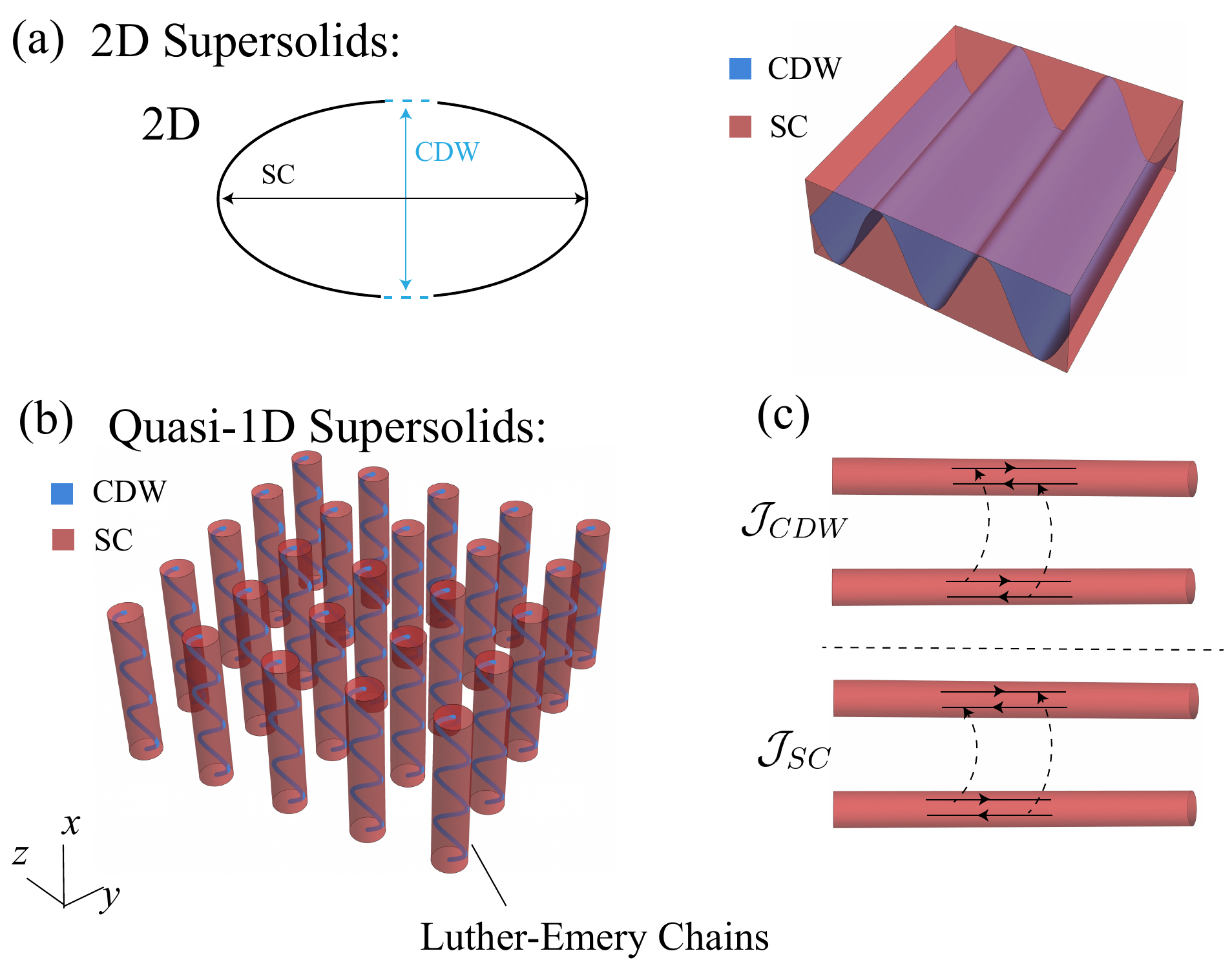}
    \caption{An illustration of supersolids arising from the coexistence of charge density waves (CDW) and superconductivity (SC). (a) 2D supersolid arising from the CDW and SC order gapping distinct parts of the Fermi surface. The CDW  channel opens a gap in the Fermi surface around the $2k_F$ nesting region (blue dashed line), while the SC channel couples states of $\bm{k}$ and $-\bm{k}$ near the Fermi energy. 
  The right panel is a schematic picture to show the coexistence of CDW and SC.  (b) a schematic plot of quasi-1D supersolids arising from the coupled Luther-Emery chains. (c) is a schematical plot of the interchain interaction of CDW and SC channels.}
    \label{fig:fig1}
\end{figure}

One particular intriguing phenomenon resulting from the interplay of different symmetry-breaking orders is the formation of supersolids. These states are characterized by the coexistence of density wave (DW) and superfluid orders \cite{Boninsegni2012}. Recent experimental realizations of supersolids in cold atom systems \cite{Recati2023} and spin systems \cite{Xiang2024} have attracted considerable attention. It was known that the superconductivity (SC) appears in the residual Fermi surfaces which are not destroyed by the CDW\cite{Balseiro1979}, which thus results in coexistence in two-dimensional charge systems according to the BCS mean-field theory [see Fig.~\ref{fig:fig1}(a)].  However, such insight does not apply to quasi-one-dimensional interacting systems, where the interplay between SC and CDW  thus is far more keen.


 Over the past decades, there has been significant interest in exploring of the competition or coexistence of SC and CDW in quasi-one-dimensional interacting systems. One prominent platform for such studies is the Luther-Emery liquid \cite{Luther1974}. Luther-Emery liquids emerge from Luttinger liquids, which exhibit spin-charge separation. Unlike generic Luttinger liquids, Luther-Emery liquids have gapped spin excitations, leaving low-energy states governed solely by charge excitations. These systems are naturally described by bosonization techniques \cite{Balents1996, Congjun2003, Tsvelik2005, Giamarchi2003} and have also been numerically confirmed in ladder-type Hubbard models using Density Matrix Renormalization Group (DMRG) methods \cite{Noack1994, Noack1997, Friedel2002}. The Luther-Emery liquid framework was adopted to study high-temperature superconductors \cite{Fradkin2015, Lubensky2000, Emery2000, Kane2001, Carr2002, Kivelson2004, Eduardo2010}.

The competition between SC and CDW in Luther-Emery liquids was initially analyzed using interladder renormalization group (RG) methods \cite{Emery2000, Ashvin2001} and interchain mean-field theories \cite{Carr2002}. These studies suggested that the transition between SC and CDW is first-order at zero temperature, precluding coexistence. However, in 2010, Jaefari, Lal, and Fradkin \cite{Eduardo2010} proposed that coexistence is possible at the self-duality point $K=1$ (where $K$ 
 is the Luttinger parameter), based on a nonlinear 
$\sigma$-model in 2+1 dimensions. More recently, Ma et al.  \cite{Ma2023} conducted a detailed analysis near the self-dual point using a mean-field approximation, concluding that the coexistence of SC and CDW is energetically unfavorable. Thus, whether Luther-Emery liquids can support ground states with coexisting SC and CDW remains to be elucidated. The problem is closely related to the exploration of supersolids in strongly correlated electronic systems, which is still poorly understood.

In this work, we provide a comprehensive theory to explore the supersolid order in Luther-Emery liquids.  The system under consideration consists of coupled Luther-Emery liquid chains, with both interchain Josephson and density wave interactions taken into account [see Fig.~\ref{fig:fig1}(b) and (c)].  We begin by introducing the interchain Luther-Emergy liquid model using the bosonization formalism (Sec. II).  By employing a path integral variational approach, we derive two coupled gapped equations to determine the CDW and SC order parameters. Solving these equations allows us to identify the quasi-one-dimensional supersolid phase regime analytically, where the CDW and SC order parameters simultaneously survive within each chain; see Sec. III. Furthermore, we numerically minimize the variational free energy and find that the quasi-one-dimensional supersolids can be more energetically favorable than the non-coexist state in the short chain length limit [See Sec. IV]. When the chain length increases, we find that the quantum fluctuations would ultimately suppress the supersolid states. 

The coexistence of SC and CDW states also makes the dynamics of the supersolid state particularly intriguing. Using field theory methods, we demonstrate that the collective motion of quasi-one-dimensional supersolids under external electric fields is governed by a quasi-Goldstone mode (Sec. V), which excludes the zero modes of $q_x = 0$.  We also show that the dynamics of zero modes can be mapped to the Josephson plasma mode in these coupled Luttinger-Emery liquid chains.


 \section{Interchain Luther-Emery liquid model from basonization}

\subsection{Luther-Emery liquids}
Let us begin by briefly introducing the bosonization framework for describing  Luther-Emery liquids. In the spinful case, the spin and charge can be separated as follows
$
 \rho_{r,c}(x)=\frac{1}{\sqrt{2}}(\rho_{r,\uparrow}+\rho_{r,\downarrow}), \rho_{r,s}(x)=\frac{1}{\sqrt{2}}(\rho_{r,\uparrow}-\rho_{r,\downarrow})  
$, where $r=R/L$ labels the right and left Fermi branches,  $c$ and $s$ label the charge and spin sectors, respectively.  The density operators are defined as  
$ \rho_{r,\sigma}(x)=\frac{1}{2\pi} \partial_x\phi_{r,\sigma}$ with $\sigma=\uparrow \text{or} \downarrow$, where $\phi_{r,\sigma}$ is a bonsic field. 

The total Hamiltonian of a 1D spinful Luttinger liquid after bosonization is given by \cite{Giamarchi2003}
\begin{eqnarray}
H=&&\frac{v_c}{2\pi} \int dx [K_c(\partial_x\theta_{-,c})^2+\frac{1}{K_c}(\partial_x\theta_{+,c})^2]\nonumber\\
&&+\frac{v_s}{2\pi} \int dx [K_s(\partial_x\theta_{-,s})^2+\frac{1}{K_s}(\partial_x\theta_{+,s})^2]\nonumber\\
&&+\int dx \frac{2g_{1\perp}}{(2\pi\alpha)^2}\cos(2\theta_{+,s}). \label{Hamil1}
\end{eqnarray}
where $v_{c(s)}$ denotes the velocity,  $K_{c(s)}$ represents the Luttinger parameter of charge  (spin) sectors, respectively. The bosonic fields are 
$\theta_{\pm, c }=\frac{1}{\sqrt{2}}(\phi_{\pm,\uparrow}+\phi_{\pm,\downarrow})$ , and $\theta_{\pm, s }=\frac{1}{\sqrt{2}}(\phi_{\pm,\uparrow}-\phi_{\pm,\downarrow})$ with $\phi_{\pm,\sigma}=\frac{1}{\sqrt{2}}(\phi_{R,\sigma}\pm \phi_{L,\sigma})$. 

The last term in Hamiltonian $H$ arises from the interaction channel $g_{1\perp}\sum_{\sigma}\psi^{\dagger}_{L,\sigma}\psi^{\dagger}_{R,-\sigma}\psi_{L,-\sigma}\psi_{R,\sigma}$, where the same spin is exchanged between branches. Note that the corresponding field operator in the bosonization form is
$\psi_{r,\sigma}(x)=\frac{1}{\sqrt{2\pi\alpha}} U_{r,\sigma} e^{irk_Fx} e^{\frac{i}{2}r[(\theta_{+,c}+r\theta_{-,c})+\sigma(\theta_{+,s}+r\theta_{-,s})]}$, where $\alpha$ is a cutoff \cite{Giamarchi2003}, $k_F$ is the Fermi wavelength, and $U_{r,\sigma}$ are the Klein factors.
Due to the presence of this $\cos$  term being proportional to $g_{1 \perp}$ in $H$, the spin sector can be gapped when $g_{1 \perp} <0$, leaving only the charge sector to contribute to the low-energy degrees of freedom \cite{Luther1974, Giamarchi2003}. 

This phase is identified as the Luther-Emery liquid, with a
low-energy Hamiltonian given by
\begin{equation}
H_0=\frac{v_c}{2\pi} \int dx [K_c(\partial_x\theta_{-,c})^2+\frac{1}{K_c}(\partial_x\theta_{+,c})^2]. \label{bosonization}
\end{equation}
Here, $\theta_{+,c}$ and $\theta_{-,c}$ satisfy the commutation relation
\begin{equation}
[\theta_{+,c}(x'),\frac{1}{\pi}\partial_x\theta_{-,c}(x)]= i\delta(x'-x). \label{comm}
\end{equation} 
Actually, $\theta_{-,c}$ and $\theta_{+,c}$ can be regarded as the phase fields for SC and CDW order parameters, respectively. The CDW order parameter is given by
    \begin{equation}
\Delta_{CDW}=\psi^{\dagger}_{R\uparrow}\psi_{L\uparrow}+\psi^{\dagger}_{R\downarrow}\psi_{L\downarrow}=2e^{-i\theta_{+,c}}\cos(\theta_{+,s})\sim 2e^{-i\theta_{+,c}},
\end{equation}
where $\theta_{+,s}$ is pinned around 0 in the spin-gapped phase. 
The SC order parameter is
\begin{equation}
\Delta_{SC}=\psi_{R,\uparrow}\psi_{L,\downarrow}-\psi_{R\downarrow}\psi_{L\uparrow}=2e^{i\theta_{-,c}}\cos(\theta_{+,s})\sim 2e^{i\theta_{-,c}}.
\end{equation}
Therefore, the competition between the SC and CDW is closely related to the Heisenberg uncertainty principle, which is most keen in one dimension. Then, the coexistence of SC and CDW is forbidden in purely one-dimensional models, and the physically relevant question is whether the coexistence is possible when a weak interchain interaction is introduced still keeping the 1D nature of the system.

\subsection{Interchain coupling}

We extend the model to higher dimensions by introducing interchain coupling through interactions. Specifically, the intra-chain Hamiltonian is given by
\begin{equation}
H_0 = \frac{v}{2\pi} \sum_{j} \int dx \left[ K(\partial_x\theta_{-,j})^2 + \frac{1}{K}(\partial_x\theta_{+,j})^2 \right].
\end{equation}
Here, $j$ is a chain index. For simplicity, we neglect the charge sector label from now on. In the presence of time-reversal symmetry, the allowed interchain Josephson and density-density interactions are described as follows \cite{Kane2002}:
\begin{eqnarray}
H_{inter} &&= -\frac{\mathcal{J}_{CDW}}{2} \sum_{\braket{i,j}} \int dx \psi^{\dagger}_{L,i} \psi_{R,i} \psi^{\dagger}_{R,j} \psi_{L,j} \nonumber\\
&&-\frac{\mathcal{J}_{SC}}{2} \sum_{\braket{i,j}} \int dx \psi^{\dagger}_{R,j} \psi_{R,i} \psi^{\dagger}_{L,j} \psi_{L,i} + h.c..
\end{eqnarray}
Here, $\mathcal{J}_{CDW}$ and $\mathcal{J}_{SC}$ represent the interaction strengths of the CDW and SC channels, respectively, and $\braket{i,j}$ denotes summation over nearest-neighbor chains. For simplicity, we have neglected the hopping $t_\perp$ of single electron between the chains, which becomes an irrelevant operator due to the spin gap (while the second order processes in $t_\perp$ can be relevant \cite{Giamarchi2003}). After bosonization, the interchain Hamiltonian becomes
\begin{eqnarray}
H_{inter} &&= -\mathcal{J}_{CDW} \int dx \sum_{\braket{i,j}} \cos(\theta_{+,i} - \theta_{+,j}) \nonumber\\
&&- \mathcal{J}_{SC} \int dx \sum_{\braket{i,j}} \cos(\theta_{-,i} - \theta_{-,j}).
\end{eqnarray}

The total action is therefore
\begin{eqnarray}
\mathcal{S} &&= \int dx d\tau \bigg[ i \sum_{j} \partial_{x} \theta_{+,j} \partial_{\tau}\theta_{-,j} + \frac{v}{2\pi} \sum_{j} \bigg(K(\partial_x \theta_{-,j})^2 \nonumber\\
&&+ \frac{1}{K} (\partial_x \theta_{+,j})^2\bigg) - \mathcal{J}_{CDW} \sum_{\braket{i,j}} \cos(\theta_{+,i} - \theta_{+,j}) \nonumber\\
&&- \mathcal{J}_{SC} \sum_{\braket{i,j}} \cos(\theta_{-,i} - \theta_{-,j}) \bigg].  \label{eq_action}
\end{eqnarray}
It can be seen that the CDW and SC degrees of freedom are coupled via the Berry phase term $i \partial_{x} \theta_{+} \partial_{\tau} \theta_{-}$ \cite{nagaosa1999, Altland2010}. In the following, we investigate the competition between CDW and SC using the variational method, without involving interchain mean-field approximations \cite{Carr2002, Ma2023}.

\section{Path Integral Variational Analysis}
\subsection{Varitional principle}
We now introduce the variational principle in the path integral formalism introduced by Feynman \cite{feynman2010}.  
For a general action $S$, the partition function $\mathcal{Z}=\int \mathcal{D}[\theta] e^{-S}$ may not be analytically solvable. In such cases, we define a variational action $S_0$ and rewrite the partition function as 
\begin{equation}
\mathcal{Z}=\int \mathcal{D}[\theta] e^{-S}=\int \mathcal{D}[\theta] e^{-S_0} e^{-(S-S_0)}=\mathcal{Z}_0\braket{e^{-(S-S_0)}}, 
\end{equation}
where $\mathcal{Z}_0=\int \mathcal{D}[\theta] e^{-S_0}$.  Recalling that the free energy is given by $F = -T \log \mathcal{Z}$ and using the inequality $\braket{e^{-(S-S_0)}} \geq e^{-\braket{(S-S_0)}}$, we obtain  
\begin{equation}
F \leq F_{\text{var}} = F_0 + T \braket{S - S_0}_0, \label{var}
\end{equation}
where $F_0 = -T \log \mathcal{Z}_0$, and $F_{\text{var}}$ denotes the varitional free energy.  

The aim of the variational method is to identify a solvable action $S_0$ that minimizes $F_{\text{var}}$. In the following, we apply this variational method to analyze the CDW and SC states described by the action $\mathcal{S}$ given in Eq.~\eqref{eq_action}.

\subsection{Varitional analysis for the  Luther-Emery liquid chains}
  For convenience, we transform the action into momentum and frequency space using \cite{Giamarchi2003, Haldane_1981}
\begin{eqnarray}
   &&\theta_{+}(\bm{r})=\frac{-(N_R+N_L)\pi x}{L}+\frac{1}{\sqrt{\beta\Omega}} \sum_{\bm{q}}\theta_{+}(\bm{q})e^{-i\bm{q}\cdot \bm{r}}\nonumber\\
&&\theta_{-}(\bm{r})=\frac{(N_R-N_L)\pi x}{L}+\frac{1}{\sqrt{\beta\Omega}} \sum_{\bm{q}}\theta_{-}(\bm{q})e^{-i\bm{q}\cdot \bm{r}},\nonumber\\
\label{theta}
\end{eqnarray}
where $L$ is the length of each chain, $\beta=1/T$, $\Omega$ is the sample area, $\bm{r}=(\bm{x},v\tau)$ and $\bm{q}=(\bm{k},\omega_n/v)$, with $\bm{x}=(x,y,z)$ and $\bm{k}=(k_x,k_y,k_z)$, $N_{r}$ is the particle number of $r$-branch.  Note that periodic boundary conditions are assumed along the $y$- and $z$-direction.  It is important to note that $k_x=n \pi/L$, and $n$ are non-zero integers here. In other words, the $\sum_{\bm{q}}$ does not include the mode $k_x=0$.  The information zero mode $k_x=0$ is encoded in the first term of $\theta_{\pm}(\bm{r})$, where $\partial_x\theta_{+}(\bm{r})$ and  $\partial_x\theta_{-}(\bm{r})$ are related to the particle number summation and difference of left and right branch, respectively.
Moreover, due to the commutation relation Eq.~\eqref{comm},  $\theta_{+}(\bm{k})$ and  $\theta_{-}(\bm{k})$ are not commute and respects [see Appendix A]
\begin{equation}
[\theta_{+}(\bm{k}'), \theta_{-}(-\bm{k})]=\frac{\pi}{k_x} \delta_{\bm{k}',\bm{k}}.
\end{equation}

In the momentum space, the quadratic term in action $\mathcal{S}$ can be conveniently rewritten as 
\begin{eqnarray}
    S&&=\sum_{\bm{q}} \Big[ ik_x\omega_n \theta_{+}(\bm{q}) \theta_{-}(-\bm{q}) 
    + \frac{vK}{2\pi} \sum_{\bm{q}} k_x^2 \theta_{-}(\bm{q}) \theta_{-}(-\bm{q}) \nonumber\\
    &&+ \frac{v}{2\pi K} \sum_{\bm{q}} k_x^2 \theta_{+}(\bm{q}) \theta_{+}(-\bm{q}) \Big] 
    - \mathcal{J}_{CDW}\int dx \int d\tau \sum_{\braket{i,j}} \nonumber\\
    && \cos(\theta_{+,i}-\theta_{+,j}) 
    - \mathcal{J}_{SC} \int dx \int d\tau \sum_{\braket{i,j}} \cos(\theta_{-,i}-\theta_{-,j}).\nonumber
\end{eqnarray}
Note that we have dropped a constant term :  $\frac{\pi\beta\Omega v}{2}(\frac{K(N_R-N_L)^2}{L^2}+\frac{(N_R+N_L)^2}{KL^2})$. 
Let us define the variational action as
\begin{equation}
    S_0=\frac{1}{2}\sum_{\bm{q}}\begin{pmatrix}
        \theta_{+}(-\bm{q})&\theta_{-}(-\bm{q})
    \end{pmatrix}G^{-1}(\bm{q})\begin{pmatrix}
        \theta_{+}(\bm{q})\\
        \theta_{-}(\bm{q})
    \end{pmatrix}.
\end{equation}
As a result, the partition function given by the variational principle becomes
\begin{equation}
    \mathcal{Z}_0=\int\mathcal{D}[\theta_{+},\theta_{-}] e^{-S_0} = \text{Det}(G).
\end{equation}
The corresponding free energy is then given by
\begin{equation}
       F_0=-T\sum_{\bm{q}}\text{tr}\{\text{log}[G(\bm{q})]\} = -T\sum_{\bm{q}}\log(\text{Det}(G)).
\end{equation}
On the other hand, the expectation value of the action $\mathcal{S}$ is
\begin{eqnarray}
    \braket{S}_0 &&= \sum_{\bm{q}}\left\{ik_x\omega_n\braket{\theta_{+}(\bm{q})\theta_{-}(-\bm{q})}_0 + \frac{vK}{2\pi}\sum_{\bm{q}} k_x^2 \times \right. \nonumber\\
    && \left.\braket{\theta_{-}(\bm{q})\theta_{-}(-\bm{q})}_0 + \frac{v}{2\pi K}\sum_{\bm{q}} k_x^2 \braket{\theta_{+}(\bm{q}) \theta_{+}(-\bm{q})}_0 \right\} \nonumber\\
    &&- \int dx\,d\tau \mathcal{J}_{CDW}\sum_{\braket{i,j}} \braket{\cos(\theta_{+,i}-\theta_{+,j})}_0 - \int dx\,d\tau \mathcal{J}_{SC} \nonumber\\
    &&\times \sum_{\braket{i,j}} \braket{\cos(\theta_{-,i}-\theta_{-,j})}_0.
\end{eqnarray}
After performing the Gaussian integration, we obtain
\begin{eqnarray}
\braket{S}_0 &&= \sum_{\bm{q}}\left(ik_x\omega_n G_{12}(\bm{q}) + \frac{vK k_x^2}{2\pi} G_{22}(\bm{q}) + \frac{vk_x^2}{2\pi K} G_{11}(\bm{q})\right) \nonumber\\
&&- \mathcal{J}_{CDW}\beta\Omega  (e^{-\frac{1}{\beta\Omega} \sum_{\bm{q}} G_{11}(\bm{q})(1-\cos k_y)}+ (k_y\leftrightarrow k_z))\nonumber\\
&&- \mathcal{J}_{SC} \beta \Omega (e^{-\frac{1}{\beta\Omega} \sum_{\bm{q}} G_{22}(\bm{q})(1-\cos k_y)}+(k_y\leftrightarrow k_z)),
\end{eqnarray}
Here, the Gaussian integral property $\braket{\cos \theta} = e^{-\frac{1}{2}\braket{\theta^2}}$ is adopted. This average has captured the fluctuation effects of phase fields.

Now, we can obtain the variational free energy through 
$
F_{\text{var}} = F_0 + T\left(\braket{S}_0 - \braket{S_0}_0\right)$, i.e., Eq.~\eqref{var}. 
Note that $ \braket{S_0}_0 = \frac{1}{2}\sum_{\bm{q}}\text{tr}[G G^{-1}] = \text{const}$ is not essential in the variational process. The variational equations are then
\begin{eqnarray}
\frac{\partial F_{\text{var}}}{\partial G_{11}(\bm{q})} = 0, \nonumber \\
\frac{\partial F_{\text{var}}}{\partial G_{22}(\bm{q})} = 0, \\
\frac{\partial F_{\text{var}}}{\partial G_{12}(\bm{q})} = 0. \nonumber
\end{eqnarray}
We obtain the following results:
\begin{eqnarray}
&&(G(\bm{q})^{-1})_{11} = \frac{vk_x^2}{2\pi K} +  [\mathcal{J}_{CDW}(k_y) e^{-\frac{1}{\beta\Omega} \sum_{\bm{q}} G_{11}(\bm{q})(1-\cos k_y)}\nonumber
\\
&&+(k_y \leftrightarrow k_z)], \label{Eq_20} \\
&&(G(\bm{q})^{-1})_{22} = \frac{vK k_x^2}{2\pi } + [\mathcal{J}_{SC}(k_y) e^{-\frac{1}{\beta\Omega} \sum_{\bm{q}} G_{22}(\bm{q})(1-\cos k_y)}, \nonumber \\
&&+(k_y \leftrightarrow k_z)], \label{Eq_21}\\
&&((G(\bm{q}))^{-1})_{12} = ik_x\omega_n. 
\end{eqnarray} 
Here, $ \mathcal{J}_{CDW/SC}(k_y) = \mathcal{J}_{CDW/SC}(1 - \cos k_y)$. Based on the form of $G(\bm{q})^{-1}$, we adopt the following ansatz for $G^{-1}(\bm{q})$:
\begin{equation}
G^{-1}(\bm{q}) = \begin{pmatrix}
  f_1(\bm{k}) & ik_x\omega_n \\
  ik_x\omega_n & f_2(\bm{k})
\end{pmatrix}, \label{Eq_inverse}
\end{equation}
where
\begin{eqnarray}
    f_1(\bm{k}) &= \frac{1}{2\pi K} \left[ v k_x^2 + \frac{\Delta^2_{CDW}}{v} (2 - \cos k_y-\cos k_z) \right], \\
    f_2(\bm{k}) &= \frac{K}{2\pi} \left[ v k_x^2 + \frac{\Delta^2_{SC}}{v} (2 - \cos k_y-\cos k_z) \right].
\end{eqnarray}
Note that the factor $(2-\cos k_y-\cos k_z)$ indicates that the CDW or SC order does not arise from intrachain interactions but rather from nearest-neighbor interchain interactions. Consequently, the resulting SC and CDW orders are anisotropic in the $x$- and $(y,z)$-directions. Nevertheless, we would use the magnitudes of $\Delta_{CDW}$ and $\Delta_{SC}$ to characterize the presence of CDW and SC orders, respectively. Moreover, the $(3+1)$ dimension analysis can be reduced to $(2+1)$ dimension by simply setting $k_y$ or $k_z$ as zero.

Now we can simplify the exponential factor in Eqs.~\eqref{Eq_20} and \eqref{Eq_21}  by performing the Matsubara frequency summation, which  yields
\begin{equation}
    \frac{1}{\beta} \sum_{n} G(\bm{q}) = G(\bm{k}) = \begin{pmatrix}
    G_{11}(\bm{k}) & 0 \\
    0 & G_{22}(\bm{k})
    \end{pmatrix}, \label{Eq_gree}
\end{equation}
where
\begin{eqnarray}
    G_{11}(\bm{k}) &= \frac{K}{2|k_x|} \sqrt{\frac{v k_x^2 + \frac{\Delta^2_{SC}}{v} (2 - \cos k_y-\cos k_z)}{v k_x^2 + \frac{\Delta_{CDW}^2}{v}(2 - \cos k_y-\cos k_z)}}, \label{Eq_26} \\
    G_{22}(\bm{k}) &= \frac{1}{2K |k_x|} \sqrt{\frac{v k_x^2 + \frac{\Delta^2_{CDW}}{v} (2 - \cos k_y-\cos k_z)}{v k_x^2 + \frac{\Delta_{SC}^2}{v}(2 - \cos k_y-\cos k_z)}}. \label{Green2}
\end{eqnarray}
Here, we have replaced $\frac{1}{\beta} \sum_{n}$ with $\int_{-\infty}^{+\infty} \frac{d\omega}{2\pi}$ by considering the low-temperature limit. The off-diagonal term in Eq.~\eqref{Eq_gree} is zero because the off-diagonal elements, as given in Eq.~\eqref{Eq_inverse}, are odd in frequency.

By comparing Eqs.~(\ref{Eq_20}-\ref{Eq_21}) with Eq.~\eqref{Eq_inverse}, the self-consistent gap equations can be written as
\begin{eqnarray}
\frac{\Delta_{CDW}^2}{2\pi Kv} &= \mathcal{J}_{CDW} e^{-\frac{1}{\Omega}\sum_{\bm{k}} G_{11}(\bm{k})(1 - \cos k_y)}, \label{Eq_28}\\
\frac{ K \Delta_{SC}^2}{2\pi v} &= \mathcal{J}_{SC} e^{-\frac{1}{\Omega}\sum_{\bm{k}} G_{22}(\bm{k})(1 - \cos k_y)}. \label{Eq_29}
\end{eqnarray}
Note that the $k_y$ and $k_z$ on the right-hand side of above equations can be exchanged ( $\sum_{\bm{k}} G_{11}(\bm{k})(1 - \cos k_y)= \sum_{\bm{k}} G_{11}(\bm{k})(1 - \cos k_z)$) and the results remain the same. 
Next, we analyze the possible supersolid phases based on the above gap equations.

\subsection{Analytical solutions for supersolids}

We now examine the self-consistent gap equation  Eq.~\eqref{Eq_28}. Inserting Eq.~\eqref{Eq_26}, the exponential factor in Eq.~\eqref{Eq_28} becomes
\begin{eqnarray}
&&\frac{1}{\Omega}\sum_{\bm{k}} G_{11}(\bm{k})(1-\cos k_y) = \frac{K}{(2\pi)^3}\int_{\epsilon}^{\Lambda} d k_x \int_{-\pi}^{\pi} dk_y \int_{-\pi}^{\pi} dk_z  \nonumber\\
&&\times \frac{1}{k_x} \sqrt{\frac{v k_x^2 + \frac{\Delta^2_{SC}}{v} (2 - \cos k_y-\cos k_z)}{v k_x^2 + \frac{\Delta_{CDW}^2}{v}(2 - \cos k_y-\cos k_z)}}(1 - \cos k_y).\nonumber\\ \label{Eq_133}
\end{eqnarray}
The cutoffs, $\epsilon$, and $\Lambda$, are introduced because the integral exhibits ultraviolet and infrared divergence with respect to $k_x$.  The upper cutoff, $\Lambda= \pi/a$, is associated with the finite lattice spacing, while the lower cutoff, $\epsilon =\pi/(L a)$, is limited by the system size along the $x$-direction. As we mentioned, the zero mode $k_x=0$ is not included in the summation. 

Unfortunately, aside from the small $v$ limit, we do not find a generic analytical solution for the integral in Eq.~\eqref{Eq_133}. In the small $v$ limit ($v\Lambda \ll \Delta_{SC}, \Delta_{CDW}$), we can approximate the sum as
\begin{eqnarray}
&&\frac{1}{\Omega}\sum_{\bm{k}} G_{11}(\bm{k})(1 - \cos k_y) \nonumber\\
&&\approx \frac{K \Delta_{SC}}{(2\pi)^2 \Delta_{CDW}}\int_{\epsilon}^{\Lambda} d k_x \frac{1}{k_x} \int_{-\pi}^{\pi} dk_y  (1 - \cos k_y) \nonumber\\
&&= \frac{K}{2\pi} \log\left(\frac{\Lambda}{\epsilon}\right) \frac{\Delta_{SC}}{\Delta_{CDW}}. \label{diver}
\end{eqnarray}
Thus, the self-consistent equation in the small $v$ limit becomes
\begin{equation}
\frac{\Delta_{CDW}^2}{2\pi Kv} = \mathcal{J}_{CDW} e^{-\frac{K}{2\pi} \log\left(\frac{\Lambda}{\epsilon}\right) \frac{\Delta_{SC}}{\Delta_{CDW}}}. \label{Eq_32}
\end{equation}
Similarly, from Eq.~\eqref{Eq_29}, we obtain the other self-consistent equation in the small $v$ limit:
\begin{equation}
\frac{K \Delta_{SC}^2}{2\pi v} = \mathcal{J}_{SC} e^{-\frac{1}{2\pi K} \log\left(\frac{\Lambda}{\epsilon}\right) \frac{\Delta_{CDW}}{\Delta_{SC}}}.
\end{equation}

Evidently, these two nonlinear equations would yield two conventional solutions:
\begin{equation}
    \Delta_{SC} = 0, \quad \Delta_{CDW} = \sqrt{2\pi Kv \mathcal{J}_{CDW}}. \label{Eq_34}
\end{equation}
and
\begin{equation}
    \Delta_{CDW} = 0, \quad \Delta_{SC} = \sqrt{2\pi v \mathcal{J}_{SC}/K}. \label{Eq_35}
\end{equation}

When assuming that both $\Delta_{SC}$ and $\Delta_{CDW}$ are non-zero, we find another solution:
\begin{eqnarray}
\Delta_{SC} &= \sqrt{2\pi v \mathcal{J}_{SC}/K}e^{-\frac{\lambda x}{2}}, \label{SC} \\
\Delta_{CDW} &= \sqrt{2\pi v K \mathcal{J}_{CDW}} e^{-\frac{\lambda}{2x}}, \label{CDW}
\end{eqnarray}
where $x = \frac{\Delta_{CDW}}{K \Delta_{SC}}$, $\lambda = \frac{\log\left(\frac{\Lambda}{\epsilon}\right)}{2\pi}$, and the value of $x$ is determined by the transcendental equation
\[
x^2 = \frac{\mathcal{J}_{CDW}}{\mathcal{J}_{SC}} e^{-\lambda (x^{-1} - x)}.
\]
Eqs.~(\ref{SC}) and \eqref{CDW} represent a charge supersolid solution, where both the CDW and SC order parameters are finite. At the self-duality point ($\mathcal{J}_{CDW} = \mathcal{J}_{SC} = \mathcal{J}$ and $K = 1$), we can analytically find
\begin{equation}
 \Delta_{SC} = \Delta_{CDW} = \sqrt{2\pi v \mathcal{J}} \left(\frac{\epsilon}{\Lambda}\right)^{\frac{1}{4\pi}} \propto (\frac{1}{L})^{\frac{1}{4\pi}}. \label{self}
\end{equation} 
Interestingly, the above form implies that the supersolid phase would be more favorable with a shorter chain length $L$.

\section{Supersolids from minimizing variational free energy}

\begin{figure*}
    \centering
    \includegraphics[width=1\linewidth]{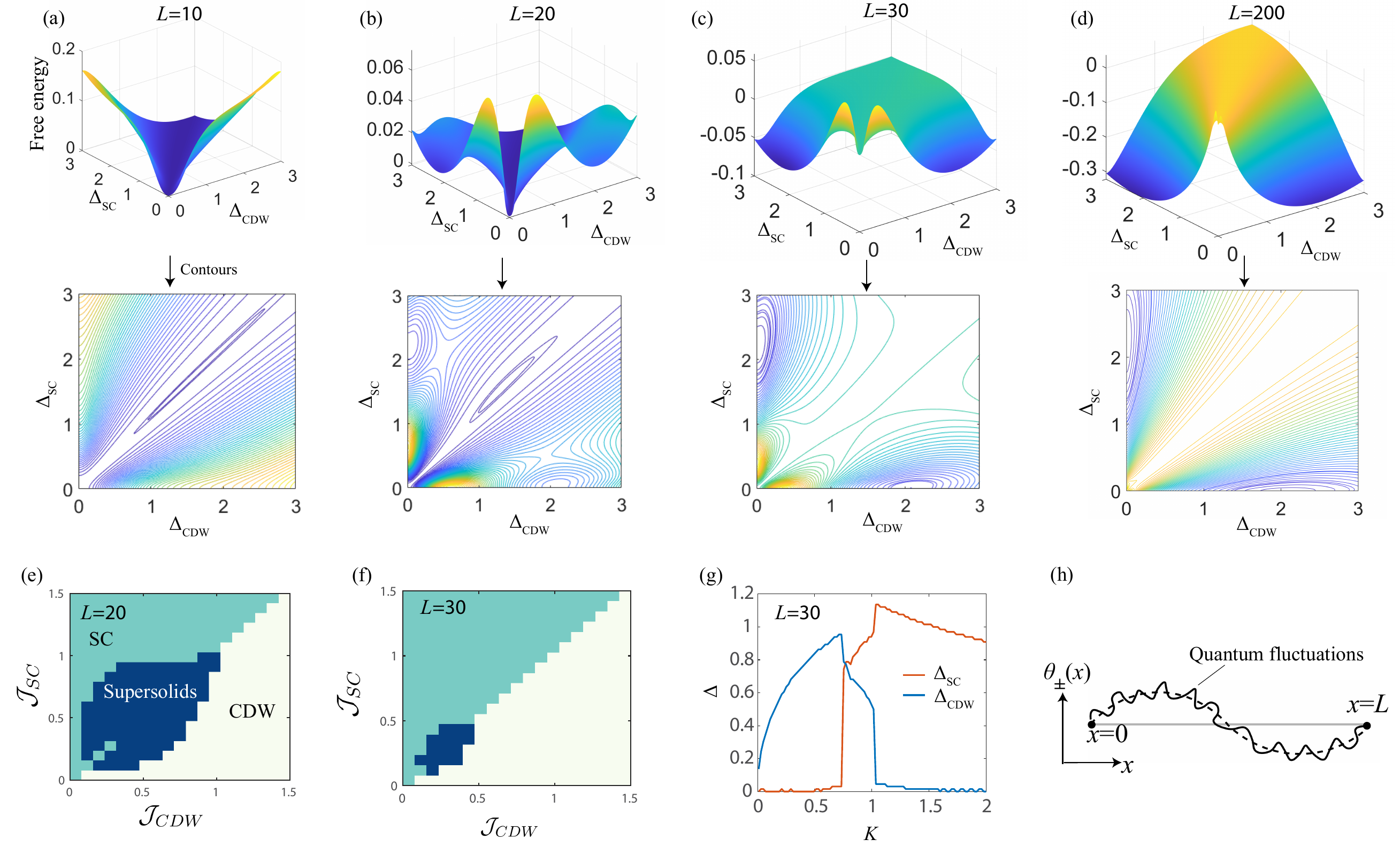}
    \caption{Supersolids from minimizing variational free energy in zero temperature limit. (a) to (d) The $\Delta_{CDW}$  and $\Delta_{SC}$ dependence of  variational free energy $\tilde{F}_{var}$ for different chain length $L=10, 20, 30, 200$, respectively. The corresponding energy contours are displayed in the lower panel. Here, $v=1$, $K=0.95$, $\mathcal{J}_{CDW}=1$, $\mathcal{J}_{SC}=1.05$. (e) and (f) The phase dependence on the interacting strength $\mathcal{J}_{CDW}$ and $\mathcal{J}_{SC}$ of $L=20$  and $L=30$, respectively. (g) The order parameter $\Delta_{CDW}$ and $\Delta_{SC}$ versus the Luttinger parameter $K$ with $L=30$, $\mathcal{J}_{CDW}=0.3$, $\mathcal{J}_{SC}=0.35$. (h) A schematical plot of low-energy phase mode with quantum fluctuations.  }
    \label{fig:fig2}
\end{figure*}

The next natural question is whether the supersolid solution we identified represents a true ground state or simply a metastable state. To address this, we directly minimize the variational free energy. As explicitly shown in Appendix C, the variational free energy after the Matsubara summation becomes
\begin{eqnarray}
    F_{\text{var}}&&\approx \frac{2}{\beta} \sum_{\bm{k}} \log (1-e^{-\beta \xi(\bm{k})})+ \sum_{\bm{k}} [\frac{vK k_x^2}{2\pi} G_{22}(\bm{k})\nonumber\\
&&+\frac{v k_x^2}{2\pi K} G_{11}(\bm{k})]-\mathcal{J}_{CDW}\Omega[e^{-\frac{1}{\Omega}\sum_{\bm{k}} G_{11}(\bm{k})(1-\cos k_y)}\nonumber\\
&&+(k_y\leftrightarrow k_z)]-\mathcal{J}_{SC}\Omega[e^{-\frac{1}{\Omega}\sum_{\bm{k}} G_{22}(\bm{k})(1-\cos k_y)} \nonumber\\
&&+(k_y\leftrightarrow k_z)].
\end{eqnarray}
Here, the quasi-particle excitation energy $\xi(\bm{k})=|k_x|^{-1}\sqrt{f_1(\bm{k})f_2(\bm{k})}$. Note that the momentum $k_x=n\pi/L$ with $n=\pm 1, \dots, \pm L$.  At zero temperature limit, it is straightforward to verify that the gap equations Eqs.\eqref{Eq_28} and \eqref{Eq_29} are given by the equations $\frac{\delta F_{\text{var}}}{\delta \Delta_{CDW}}=0$ and $\frac{\delta F_{\text{var}}}{\delta \Delta_{SC}}=0$ (see Appendix C for details).  For simplicity, we show the results of the 
(2+1)-dimensional case in the following presentation of this section ( set $k_z=0$).   However, we have verified that the free energy landscape in the 
(3+1)-dimensional case exhibits similar features to the 
(2+1)-dimensional case (see Appendix D). Therefore, the conclusions presented below also apply to the 
(3+1)-dimensional scenario.

Figure 2(a) to 2(d) illustrate the dependence of the variational free energy on $\Delta_{CDW}$ and $\Delta_{SC}$ for various chain lengths $L$, with parameters $K = 0.95$, $\mathcal{J}_{CDW} = 1$, and $\mathcal{J}_{SC} = 1.05$. The lower panels present the corresponding energy contours for each free energy landscape. Notably, the supersolid states gradually emerge as the global minima in the free energy landscape as the chain length decreases. This observation aligns with the implications of Eq.\eqref{self}. By tuning the interaction strengths and minimizing the free energy, the phase diagrams in Fig.\ref{fig:fig2}(e) and Fig.~\ref{fig:fig2}(f) are obtained. As anticipated, the supersolid phase becomes more favorable when $\mathcal{J}_{CDW}$ and $\mathcal{J}_{SC}$ are comparable, and the extent of the supersolid regime increases with decreasing $L$.
Figure~\ref{fig:fig2}(g) depicts the $K$ dependence of the order parameters $\Delta_{CDW}$ and $\Delta_{SC}$. It is evident that there exists a range of $K$ values where both $\Delta_{SC}$ and $\Delta_{CDW}$ are finite. Therefore, based on the variational analysis, we conclude that the supersolid states gradually behave as true ground states in our model in the short-chain limit.

The suppression of supersolid states in the long-chain limit can be attributed to quantum fluctuations. As previously noted, the CDW and SC phase fields do not commute. Consequently, the coexistence of these phase fields in real space leads to enhanced quantum fluctuations, as illustrated in Fig.~\ref{fig:fig2}(h).
The soft mode (small $k_x$) plays a dominant role in the low-energy physics. As the chain length increases, quantum fluctuations associated with the soft mode eventually destabilize the supersolid states.
Formally, the soft modes induce an infrared divergence (see Eq.~\eqref{diver}), which ultimately destroys the long-range supersolid order. This scenario closely resembles the Mermin–Wagner theorem \cite{Mermin1966}, which prohibits long-range order breaking of continuous symmetries in dimensions $d \leq 2$.

It is worth noting that we have assumed the periodic boundary condition along the chain (each chain effectively forms a ring).   In the case of open boundary conditions,  there are conventional $2k_F$ Friedel oscillations at the edge.   As a result, the open chain cannot be too short to ensure that the density oscillations induced by Friedel oscillations can be distinguished from those caused by the supersolid states.

\section{Collective motion of the CDW plus SC state}
In this section, we investigate the collective motion of states combining CDW and SC orders. In conventional incommensurate density wave systems, such as spin density waves and charge density waves, collective motion can occur under critical external currents once impurity pinning effects are overcome. The low-energy mode governing this collective motion is known as the phason. Since supersolids also exhibit density wave characteristics, the phason is likewise relevant for incommensurate supersolids. Moreover, the dynamics of supersolids are particularly intriguing due to the additional presence of superconducting features. 
 \subsection{Quasi-Goldstone mode}
 Under a static electric field $E=-\nabla \varphi$, the dynamics of such supersolid states can be captured by the following Lagrangian
\begin{equation}
\mathcal{L}= \mathcal{L}_{EM}+\mathcal{L}_{int}+ \mathcal{L}_{phason},
\end{equation}
where the electromagnetic potential term 
 $\mathcal{L}_{EM}=-\frac{1}{8\pi} (\nabla  \varphi)^2$, and the interaction between electric field and phason 
$\mathcal{L}_{int} =i 
 e\varphi(x) \partial_x \theta_{+}(x)$. 
The Lagrangian form of phason $ \mathcal{L}_{phason}$ is deduced from expanding Eq.~\eqref{Eq_inverse} with a small momentum $\bm{q}=(q_x,q_{\perp})$ with $q_{\perp}=(q_y, q_z)$. Again, the zero mode $q_x=0$ is not included there, and the corresponding dynamics will be studied in the later section.  Eventually, the Lagrangian $\mathcal{L}$ in the momentum space can be written as 
\begin{eqnarray}
\mathcal{L}&&=\frac{q^2}{8\pi} \varphi(\bm{q})\varphi(-\bm{q})+\frac{1}{2} e q_x(-\varphi(\bm{q})\theta_{+}(-\bm{q})+ \varphi(-\bm{q})\theta_{+}(\bm{q}))\nonumber\\
&&+ \frac{1}{2m}(\rho_{s\parallel} q_x^2+ \rho_{s\perp} q_{\perp}^2) \theta_{-}(\bm{q}) \theta_{-}(-\bm{q})\nonumber\\
&&+ \frac{1}{2m}(\rho_{c\parallel} q_x^2+ \rho_{c\perp} q_{\perp}^2) \theta_{+}(\bm{q}) \theta_{+}(-\bm{q})\nonumber\\
&&+\frac{i\omega_n q_x}{2}(\theta_{+}(\bm{q})\theta_{-}(-\bm{q})+\theta_{-}(+\bm{q})\theta_{+}(\bm{q})).
\end{eqnarray}
where we have defined $\frac{\rho_{c\parallel}}{m}=\frac{v}{\pi K}$, $\frac{\rho_{c\perp}}{m}=\frac{\Delta_{CDW}^2}{2\pi Kv}$, $\frac{\rho_{s\parallel}}{m}=\frac{Kv}{\pi }$, and $\frac{\rho_{s\perp}}{m}=\frac{K\Delta^2_{SC}}{2\pi v }$ with $m$ as an effective mass, and $\rho$ as density of each channel.
The total action is $S=\sum_{\bm{q}}\mathcal{L}(\bm{q})$. After integrating out the field $\varphi$, the effective action becomes
\begin{equation}
S_{eff}=\sum_{\bm{q}} \begin{pmatrix}
    \theta_{+}(\bm{q}) &\theta_{-}(\bm{q})
\end{pmatrix} G'^{-1}(\bm{q})\begin{pmatrix}
    \theta_{+}(-\bm{q})\\
    \theta_{-}(-\bm{q}) 
\end{pmatrix}.
\end{equation}
Here,
\begin{equation}
G'^{-1}(\bm{q}) =\begin{pmatrix}
    \frac{2\pi e^2 q_x^2}{q^2}+\frac{\rho_{c\parallel} q_x^2+ \rho_{c\perp} q_{\perp}^2}{2m}  & \frac{i\omega_n q_x}{2}\\
   \frac{i\omega_n q_x}{2} &  \frac{\rho_{s\parallel} q_x^2+ \rho_{s\perp} q_{\perp}^2}{2m} 
\end{pmatrix}.
\end{equation}
Therefore, we find the electric field can excite a collective mode for this quasi-1D supersolids as:
\begin{equation}
\frac{\omega^2_{q} q_x^2}{4}= ( \frac{2\pi e^2 q_x^2}{q^2}+\frac{\rho_{c\parallel} q_x^2+ \rho_{c\perp} q_{\perp}^2}{2m}) (\frac{\rho_{s\parallel} q_x^2+ \rho_{s\perp} q_{\perp}^2}{2m}).
\end{equation}
When the electric field-driven phase mode in the CDW channel is dominant,  the dispersion of the mode is approximately: \begin{equation}
\omega_{q}^2\approx \frac{4\pi e^2}{m q^2}(\rho_{s\parallel} q_x^2+ \rho_{s\perp} q_{\perp}^2). 
\end{equation}
The above dispersion characterizes a collective mode for the CDW plus SC states. 

This collective mode stems from the break of translational symmetry in supersolid states, which represents as a Goldstone mode.   However, unlike conventional Goldstone, the zero-mode $q_x=0,q_y, q_z\neq 0$ does not contain in the above treatment.  In this sense, we shall call such a mode as quasi-Goldstone mode.

\subsection{Zero mode dynamics: Josephson plasma mode }

As mentioned earlier, the zero modes ($q_x=0$, $q_y, q_z\neq 0$ ) shall be treated separately in finite chains. Such modes can be excited when currents are applied perpendicular to the chains.     According to Eqs.~\eqref{eq_action} and \eqref{theta}, can be described by the following action
\begin{eqnarray}
&&\mathcal{S}_{zero} \simeq \int dt [-\pi N \sum_{j} \partial_t \theta_{-,j}+ \frac{v K}{2\pi} \sum_{j} \int (\partial_x \theta_{-,j})^2 \nonumber\\
&&- \mathcal{J}_{SC} \sum_{\braket{i,j}} \cos(\theta_{-,i} -  \theta_{-,j})+\frac{1}{2}\sum_{i\neq j} \frac{e^2 N_i N_j}{|r_i-r_j|}].
\end{eqnarray}
where $N_j=N_{L,j}+N_{R,j}$, we used $\theta_{+,j}=- N_{j} \pi x/L$, and constant terms are omitted as they do not affect the dynamical properties. 
 The action $S$ resembles that of a Josephson array. The first term is a Berry phase term, the second term represents the phase stiffness energy within each array, and the third term corresponds to the Josephson coupling energy between nearest-neighbor arrays. For zero mode, the second term can be dropped. We also have replaced the Coulomb interaction with a long-range density interaction (the last term). As the charge distribution for the zero modes is uniform within each chain and does not fluctuate, the long-range Coulomb interaction can not be negligible.

In the continuum limit, the action can be rewritten as:
\begin{eqnarray}
&&\mathcal{S}_{zero} \approx \int dt[\int d r_{\perp}  (-\pi N (r_{\perp}, t) \partial_t\theta_{-}+\frac{\mathcal{J}_{SC}}{2} (\partial_{r_{\perp}}\theta_{-})^2)\nonumber\\
&&+\frac{1}{2} \int dr_{\perp} dr_{\perp}' \frac{e^2 N(r_{\perp},t) N(r_{\perp}',t)}{|r_{\perp}-r_{\perp}'|}].
\end{eqnarray}
Using the Euler-Lagrange equations, the equation of motion for $\theta_{-}$ in the zero-mode state is given by
\begin{equation}
     \pi \partial_t N(r_{\perp},t)=J_{SC} \partial_{r_{\perp}}^2 \theta_{-}.\label{Eq_49}
\end{equation}
The left-hand side represents the charging effects, while the right-hand side describes the phase-gradient energy. Similarly, by varying the action with respect to  $N(r_{\perp}, t)$,  we obtain
\begin{equation}
\pi\partial_t\theta_{-}= e^2 \int d r'_{\perp}  \frac{N(r_{\perp}', t)}{|r_{\perp}-r'_{\perp}|}.\label{Eq_50}
\end{equation}
Inserting Eq.~\eqref{Eq_50} into Eq.~\eqref{Eq_49},
we find
\begin{equation}
\pi^2\partial_t^2\theta_{-}(r_{\perp},t)=e^2J_{SC}\int dr'_{\perp} \frac{ \partial^2_{r'_{\perp}}\theta_{-}(r'_{\perp},t)}{|r_{\perp}-r'_{\perp}|}.\label{Eq_plas}
\end{equation}

After a Friouer transformation (see Appendix E), we find that the Eq.~\eqref{Eq_plas} results in a Josephson plasma mode:
\begin{equation}
\omega_{q}= v_{J} \sqrt{|q_{\perp}|}.
\end{equation}
with a propagating velocity:   $v_{J}=\sqrt{\frac{2e^2 J_{SC}}{\pi}}$. The square root behavior in $\omega_{q}$ is a typical 2D plasma dispersion of long-wave limit.
Thus, we have shown that the dynamics of the zero mode in our quasi-one-dimensional supersolid states behave as a  Josephson plasma mode.

 \section{Discussion}

In summary, we have systematically investigated the supersolid states and their dynamics using a coupled Luttinger-Emery liquid chain model. Our work provides significant insights into the properties of supersolid phases in quasi-one-dimensional strongly interacting systems.


It is known that the CDW phase can emerge as an induced order parameter from the mixing of uniform SC order and pairing density wave (PDW) order \cite{Agterberg}. Actually, PDWs have also been studied in Luther-Emery liquids \cite{Soto2015, Yahui2022}. Although the coexistence of uniform superconductors and CDW does not necessarily guarantee the appearance of PDW order, in principle, PDWs can intertwine with both the CDW and uniform SC orders. The interplay between PDWs and the charge supersolid phase presents an interesting avenue for future research.

Since the Luther-Emery liquid phase was originally introduced to understand the properties of high-$T_c$ superconductors, we also discuss some potential experimental implications of our theory. One key implication is that the coexistence of SC and CDW (or supersolids) is more likely to occur in ribbon or slab systems, where the length of the Luttinger liquid chain is limited. The systems with short-range orders can also be regarded as finite systems.   For example, 
in the pseudogap phase, the orders possibly fluctuate and are in a short range; the strip phase can be local and exhibit smectic behavior \cite{Kivelson1998};  the domain walls (such as structure domain, CDW, or SC domains), also often exhibit short-range orders. Our theory may provide some insight for  the coexistence of SC and CDW in these scenarios.

 \section*{Acknowledgments}
Y.M.X.  acknowledges financial support from the RIKEN Special Postdoctoral Researcher (SPDR) Program.
	N.N. was supported by JSPS KAKENHI Grant No. 24H00197 and 24H02231.
	N.N. was also supported by the RIKEN TRIP initiative.  	
	
\begin{appendix}
\section{The  phase operator in momentum space and its relation with bosonic operators}
In this Appendix section, let us look at the phase operator in momentum space and its relation with original bosonic fields.  In terms of original bosonic operators,
\begin{eqnarray}
   &&\theta_{+}(\bm{x})=\frac{-(N_R+N_L)\pi x}{L}\nonumber\\&&-\frac{i\pi}{\sqrt{\Omega}} \sum_{ k_x\neq 0, k_y,k_z} (\frac{|k_x|}{2\pi})^{1/2}\frac{1}{k_x} e^{-i\bm{k}\cdot \bm{x}} (b^{\dagger}_{\bm{k}}+ b_{-\bm{k}}),\\
&&\theta_{-}(\bm{x})=\frac{(N_R-N_L)\pi x}{L}\nonumber\\
&&+\frac{i\pi }{\sqrt{\Omega}} \sum_{ k_x\neq 0,k_y, k_z}(\frac{|k_x|}{2\pi})^{1/2}\frac{1}{|k_x|}e^{-i\bm{k}\cdot \bm{x}} (b^{\dagger}_{\bm{k}}- b_{-\bm{k}}).
\end{eqnarray}
Here, the bosonic operator is related by the density operator by $b_{\bm{k}}=\sqrt{\frac{2\pi}{k_x L}}\rho_{-\bm{k}}$ and $b^{\dagger}_{\bm{k}}=\sqrt{\frac{2\pi}{k_x L}}\rho_{\bm{k}}$ .
In the limit $L\mapsto \infty$, 
\begin{equation}
[\theta_{+}(\bm{x}'),\partial_{x}\theta_{-}(\bm{x})]=i\pi \delta(\bm{x}-\bm{x}').
\end{equation}
Compared to the main text Eq.~\eqref{theta}, we can define
\begin{eqnarray}
\theta_{+}(\bm{k})=-i\pi (\frac{|k_x|}{2\pi})^{1/2}\frac{1}{k_x} (b^{\dagger}_{\bm{k}}+ b_{-\bm{k}}),\\
\theta_{-}(\bm{k})=i\pi (\frac{|k_x|}{2\pi})^{1/2}\frac{1}{|k_x|}  (b^{\dagger}_{\bm{k}}- b_{-\bm{k}}).
\end{eqnarray}
The commutation relation between $\theta_{+}(\bm{k})$ and $\theta_{-}(\bm{k})$ are thus given by
\begin{equation}
[\theta_{+}(\bm{k}'), \theta_{-}(-\bm{k})]=\frac{\pi}{k_x} \delta_{\bm{k}, \bm{k}'}.
\end{equation}

\section{The interchain term upon Gaussian averaging}

In the main text, we deal with the following interchain term:
\begin{eqnarray}
&&\mathcal{J}_{CDW}\int_{0}^{L} dx \int_{0}^{\beta} d \tau  \sum_{\braket{i,j}}\braket{\cos(\theta_{+,i}-\theta_{+,j})}_0=\nonumber\\
&&=\mathcal{J}_{CDW} \int_{0}^{L} dx \int_{0}^{\beta} d \tau   \sum_{\braket{i,j}} e^{-\frac{1}{2}\braket{(\theta_{+,i}-\theta_{+,j})^2}}\nonumber\\
&&= \beta L  \sum_{\braket{i,j}}  \mathcal{J}_{CDW}   e^{-\frac{1}{2\beta L} \sum_{n, k_x\neq 0} (\tilde{G}_{ij}(\omega_n, k_x))_{11} } \nonumber\\
&&=\mathcal{J}_{CDW}
\beta \Omega ( e^{-\frac{1}{\beta \Omega}\sum_{n,\bm{k}_{\perp}, k_x,\neq 0}G_{11}(\omega_n,\bm{k})(1-\cos k_y)}\nonumber\\
&&+e^{-\frac{1}{\beta \Omega}\sum_{n,\bm{k}_{\perp}, k_x,\neq 0}G_{11}(\omega_n,\bm{k})(1-\cos k_z)}). 
\end{eqnarray}
Here, 
\begin{eqnarray}
    \tilde{G}_{ij}(\omega_n, k_x)&&= G_{ii}(\omega_n, k_x)+G_{jj}(\omega_n, k_x)\nonumber\\
    &&- G_{ij}(\omega_n,k_x)-G_{ji}(\omega_n, k_x).
\end{eqnarray}
and we have used
$G_{ij}(\omega_n. k_x)= \frac{1}{L_yL_z}\sum_{\bm{k}_{\perp}} G(\omega_n,\bm{k}) e^{i\bm{k}_{\perp}\cdot (\bm{r}_i-\bm{r}_j)}$.
where $\bm{k}_{\perp}=(k_x,k_y)$.

Similarly, for the SC channel,
\begin{eqnarray}
&&\mathcal{J}_{SC}\int_{0}^{L} dx \int_{0}^{\beta} d \tau  \sum_{\braket{i,j}}\braket{\cos(\theta_{-,i}-\theta_{-,j})}_0=\nonumber\\
&&\mathcal{J}_{SC}
\beta \Omega ( e^{-\frac{1}{\beta \Omega}\sum_{n,\bm{k}_{\perp}, k_x,\neq 0}G_{22}(\omega_n,\bm{k})(1-\cos k_y)}\nonumber\\&&+e^{-\frac{1}{\beta \Omega}\sum_{n,\bm{k}_{\perp}, k_x,\neq 0}G_{22}(\omega_n,\bm{k})(1-\cos k_z)}). 
\end{eqnarray}

\section{Free energy calculation and gap equations}

The free energy is 
\begin{eqnarray}
    F_0=-T\sum_{\bm{q}} \log( \text{Det}G)).
\end{eqnarray}
with
\begin{eqnarray}
    G^{-1}(\bm{q})=\begin{pmatrix}
        f_1(\bm{k})& i k_x\omega_n\\
        ik_x\omega_n& f_2(\bm{k})
    \end{pmatrix}
\end{eqnarray}
and
    \begin{eqnarray}
    f_1(\bm{k}) &= \frac{1}{2\pi K} \left[ v k_x^2 + \frac{\Delta^2_{CDW}}{v} (2 - \cos k_y-\cos k_z) \right] \label{f1}, \\
    f_2(\bm{k}) &= \frac{K}{2\pi} \left[ v k_x^2 + \frac{\Delta^2_{SC}}{v} (2 - \cos k_y-\cos k_z) \right] \label{f2}.
\end{eqnarray}
As $\text{Det}[G]=1/\text{Det}[G^{-1}]$,
\begin{equation}
    F_0=T\sum_{n,\bm{k}} \log(\beta^2(f_1(\bm{k})f_2(\bm{k})+\omega_n^2k_x^2)).
\end{equation}
Here, $\beta^2$ is introduced to make the term in the log to be dimensionless.
Then we can find
\begin{eqnarray}
 &&F_0=\frac{1}{\beta}\sum_{n,\bm{k}}[\log( \beta (|k_x|^{-1}\sqrt{f_1(\bm{k}) f_2(\bm{k}) }-i\omega_n))\nonumber\\
&&+ \log( \beta  (|k_x|^{-1}\sqrt{f_1(\bm{k}) f_2(\bm{k})}+i\omega_n))]\nonumber.
\end{eqnarray}

Using the identity    $\frac{1}{\beta}\sum_{n} \log(\beta(-i\omega_n+\xi))=\frac{1}{\beta}\log(1-e^{-\beta \xi})$ for bosons, we find
\begin{eqnarray}
    F_0&&=\frac{2}{\beta} \sum_{\bm{k}} \log\left[ 1-e^{-\beta 
\xi(\bm{k})}\right].
\end{eqnarray}
where the background with zero gap is subtracted and the excitation energy
\begin{equation}
\xi(\bm{k})=|k_x|^{-1}\sqrt{f_1(\bm{k}) f_2(\bm{k}) }.
\end{equation}

According to Eq.~\eqref{Eq_26} and \eqref{Green2}, using the form of $f_1(\bm{k})$ and $f_2(\bm{k})$, the Green's function is rewritten as
\begin{eqnarray}
    G_{11}(\bm{k})&&=\frac{1}{2|k_x|}\sqrt{\frac{f_2(\bm{k})}{f_1(\bm{k})}},\\
      G_{22}(\bm{k})&&=\frac{1}{2|k_x|}\sqrt{\frac{f_1(\bm{k})}{f_2(\bm{k})}}.
\end{eqnarray}

Using we can express the other part of the variational free energy $T\braket{S}_0$ as
\begin{eqnarray}
T\braket{S}_0&&=\sum_{\bm{k}} [\frac{vK k_x^2}{2\pi} G_{22}(\bm{k})+\frac{v k_x^2}{2\pi K} G_{11}(\bm{k})]\nonumber\\
&&-\mathcal{J}_{CDW}\Omega[e^{-\frac{1}{\Omega}\sum_{\bm{k}} G_{11}(\bm{k})(1-\cos k_y)}+(k_y\leftrightarrow k_z)]\nonumber\\
&&-\mathcal{J}_{SC}\Omega[e^{-\frac{1}{\Omega}\sum_{\bm{k}} G_{22}(\bm{k})(1-\cos k_y)} +(k_y\leftrightarrow k_z)]\nonumber\\
\end{eqnarray}
Now we can express the variational free energy $ F_{\text{var}}\approx F_0+ T\braket{S}_0$ using the $f_1(\bm{k})$ and $f_2(\bm{k})$. Note that $\braket{S_0}_0$ has been dropped there as it is a constant. 
We can use 
\begin{eqnarray}
 \frac{\delta F_{\text{var}}}{\delta \Delta_{CDW}}&&= \frac{\delta F_{\text{var}}}{\delta f_1}\frac{\delta f_1}{\delta \Delta_{CDW}}=0,\\
\frac{\delta F_{\text{var}}}{\delta \Delta_{SC}}&&= \frac{\delta F_{\text{var}}}{\delta f_2}\frac{\delta f_2}{\delta \Delta_{SC}}=0.
\end{eqnarray}
This results in the variational equations in terms of $f_1$ and $f_2$ as
\begin{eqnarray}
\frac{\delta F_{\text{var}}}{\delta f_1}=0,\\
\frac{\delta F_{\text{var}}}{\delta f_2}=0.
\end{eqnarray}
With these two variational equations, we obtain:
\begin{eqnarray}
&&\frac{1}{e^{\beta \xi}-1} \sqrt{\frac{f_2}{f_1}}+ \frac{1}{2\sqrt{f_1 f_2}} \{\frac{ v K k_x^2}{2\pi}+ J_{SC}[(1-\cos k_y) \times \nonumber\\
&&e^{-\frac{1}{\Omega}\sum_{\bm{k}} \frac{1}{2 |k_x|} \sqrt{\frac{f_1}{f_2}}(1-\cos k_y)}+ (k_y \leftrightarrow k_z)] \}-\nonumber\\
&& \frac{1}{2 f_1}\sqrt{\frac{f_2}{f_1}}\{\frac{ vk_x^2}{2\pi K}+ J_{CDW}[(1-\cos k_y) e^{-\frac{1}{\Omega} \sum_{\bm{k}} \frac{1}{2|k_x|}\sqrt{\frac{f_2}{f_1}}(1-\cos k_y)}\nonumber\\
&&+ (k_y\leftrightarrow k_z)]\},
\end{eqnarray}

\begin{eqnarray}
&&\frac{1}{e^{\beta \xi}-1} \sqrt{\frac{f_1}{f_2}}- \frac{1}{2 f_2}\sqrt{\frac{f_1}{f_2}} \{\frac{ v K k_x^2}{2\pi}+ J_{SC}[(1-\cos k_y) \times \nonumber\\
&&e^{-\frac{1}{\Omega}\sum_{\bm{k}} \frac{1}{2 |k_x|} \sqrt{\frac{f_1}{f_2}}(1-\cos k_y)}+ (k_y \leftrightarrow k_z)] \}+\nonumber\\
&& \frac{1}{2 \sqrt{f_1 f_2}}\{\frac{ vk_x^2}{2\pi K}+ J_{CDW}[(1-\cos k_y) e^{-\frac{1}{\Omega} \sum_{\bm{k}} \frac{1}{2|k_x|}\sqrt{\frac{f_2}{f_1}}(1-\cos k_y)}\nonumber\\
&&+ (k_y\leftrightarrow k_z)]\}.
\end{eqnarray}

At zero temperature limit $\beta \rightarrow \infty$, the above two equations are satisfied  when
\begin{eqnarray}
f_1(\bm{k})&&=\frac{v  k_x^2}{2\pi K}+\mathcal{J}_{CDW}[(1-\cos k_y) e^{-\frac{1}{\Omega} \sum_{\bm{k}} G_{11}(\bm{k})(1-\cos k_y)}\nonumber\\
&& +(k_y \leftrightarrow k_z)]\\
f_2(\bm{k})&&=\frac{v K k_x^2}{2\pi }+\mathcal{J}_{SC}[(1-\cos k_y) e^{-\frac{1}{\Omega} \sum_{\bm{k}} G_{22}(\bm{k})(1-\cos k_y)}\nonumber\\
&& +(k_y \leftrightarrow k_z)]
\end{eqnarray}

According to the form Eqs.~\eqref{f1} and \eqref{f2}, we obtain the gap equations as
\begin{eqnarray}
\frac{\Delta_{CDW}^2}{2\pi Kv} &= \mathcal{J}_{CDW} e^{-\frac{1}{\Omega}\sum_{\bm{k}} G_{11}(\bm{k})(1 - \cos k_y)}\\
\frac{ K \Delta_{SC}^2}{2\pi v} &= \mathcal{J}_{SC} e^{-\frac{1}{\Omega}\sum_{\bm{k}} G_{22}(\bm{k})(1 - \cos k_y)}.
\end{eqnarray}
The above equations are consistent with what we obtained in the main text.
\begin{figure}
    \centering
    \includegraphics[width=1\linewidth]{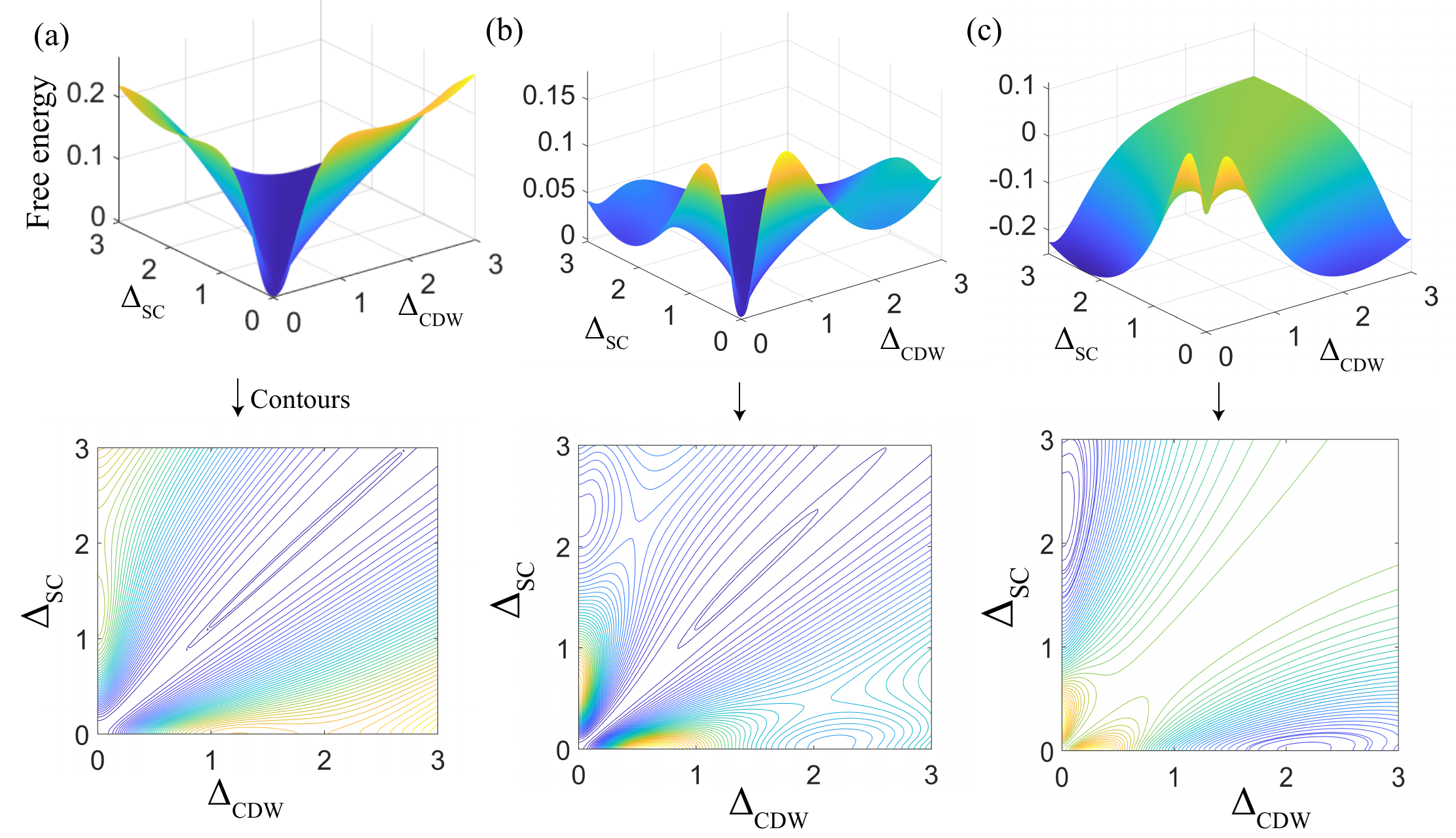}
    \caption{Free energy landscape in (3+1) dimensions. (a)–(c) The dependence of the variational free energy, $\tilde{F}_{var}$, on $\Delta_{CDW}$ and $\Delta_{SC}$, along with the corresponding energy contours, for chain lengths $L = 10, 15, 30$, respectively. The parameters are $v = 1$, $K = 0.95$, $\mathcal{J}_{CDW} = 1$, and $\mathcal{J}_{SC} = 1.05$.  }
    \label{fig:figs1}
\end{figure}
\section{Free energy landscape in (3+1) dimension}
In this Appendix section, we display the free energy landscape and the corresponding energy contours of (3+1) dimension with various chain lengths in Fig.~\ref{fig:figs1}. Overall, the features of the free energy landscape are consistent with those observed in the $(2+1)$-dimensional case shown in the main text Fig.~\ref{fig:fig2}. Consequently, the conclusions presented in Sec. IV of the main text also applies to the $(3+1)$-dimensional case.

\section{Fouriouer transformation of Coulomb potential and plasma mode}
In this section, we present how to simplify
\begin{equation}
\pi^2\partial_t^2\theta_{-}(r_{\perp},t)=e^2J_{SC}\int dr'_{\perp} \frac{ \partial^2_{r'_{\perp}}\theta_{-}(r'_{\perp},t)}{|r_{\perp}-r'_{\perp}|}.
\end{equation}
The right-hand side
\begin{eqnarray}
\int dr'_{\perp} \frac{ \partial^2_{r'_{\perp}}\theta_{-}(r'_{\perp},t)}{|r_{\perp}-r'_{\perp}|}&&= \int \frac{d q_{\perp}}{(2\pi)^2} \frac{2\pi}{|q_{\perp}|} \int dr'_{\perp}e^{-i q_{\perp} (r_{\perp}-r_{\perp}')} \nonumber\\
&&\times \partial^2_{r'_{\perp}}\theta_{-}(r'_{\perp},t)\nonumber\\
&&=-\int \frac{d q_{\perp}}{(2\pi)^2}2\pi |q_{\perp}| \theta_{-}(q_{\perp}, t).
\end{eqnarray}
Therefore, the plasma mode is given by
\begin{equation}
\pi^2\omega^2=2\pi  e^2 J_{SC}|q_{\perp}|
\end{equation}
i.e., the dispersion of the plasma mode is given by
\begin{equation}
\omega_{q}= v_{J} \sqrt{|q_{\perp}|}.
\end{equation}
where
\begin{equation}
    v_{J}=\sqrt{\frac{2e^2 J_{SC}}{\pi}}.
\end{equation}
\end{appendix}

%

\end{document}